\documentstyle[epsf,amssymb,12pt]{article}
\def\bc{\begin{center}}
\def\ec{\end{center}}
\def\be{\begin{equation}}
\def\ee{\end{equation}}
\newcommand{\nn}{\nonumber}
\newcommand{\scr}{\scriptscriptstyle }
\newcommand{\beqn}{\begin{eqnarray}}
\newcommand{\eeqn}{\end{eqnarray}}
\newcommand{\beq}{\begin{equation}}
\newcommand{\eeq}{\end{equation}}
\newcommand{\bea}{\begin{eqnarray}}
\newcommand{\eea}{\end{eqnarray}}
\newcommand{\gev}{{\rm GeV}}
\newcommand{\mev}{{\rm MeV}}

\begin{document}
\pagestyle{empty} 
\vspace{-0.6in}
\begin{flushright}
\hspace*{-5cm}
ROME1 prep. 98/1227\\
\hspace*{-5cm}
LPTHE - Orsay 98/63\\
\hspace*{-5cm}
ROME3 prep. 98/6\\
\end{flushright}
\vskip 2. cm
\centerline{\LARGE{\bf{Non-perturbatively Improved Heavy-Light }}}
\vskip 2mm
\centerline{\LARGE{\bf{ Mesons\,:\, Masses and Decay Constants}}}
\vskip 2cm
\centerline{\large \bf{D.~Becirevic$^a$, Ph.~Boucaud$^a$, J.P.~Leroy$^a$,}} 
\centerline{\large \bf{V.~Lubicz$^b$,  G.~Martinelli$^{c}$, F.~Mescia$^c$, 
F.~Rapuano$^c$}}
\begin{center}
{\sl $^a$ Laboratoire de Physique Th\'eorique et Hautes Energies{\footnote{Laboratoire associ\'e au Centre National de la Recherche Scientifique - URA D00063.}},\\
Universit\'e de Paris XI, B\^atiment 211, 91405 Orsay Cedex, France.\\
$^b$ Dip. di Fisica, Univ. di Roma Tre and
INFN, Sezione di Roma Tre,\\ Via della Vasca Navale 84, I-00146
 Rome, Italy.\\
$^c$ Dip. di Fisica, Univ. ``La Sapienza"  and
INFN, Sezione di Roma,\\ P.le A. Moro, I-00185 Rome, Italy.}\\
\end{center}
\vskip 12mm
\abstract{We present a study of the heavy-light spectrum and of the $D$- and $B$-meson decay constants. The results were obtained in the quenched approximation, by using the non-perturbatively improved Clover  lattice action
at $\beta=6.2$, with a sample of $100$ configurations, on a $24^3 \times 64$ lattice. After a careful analysis of the
systematic errors present in the extraction of the physical results,
by assuming quite conservative discretization errors,  we find
$f_{D_s}=231 \pm 12^{+6}_{-1} $~MeV, 
$f_{D} = 211 \pm 14^{+0}_{-12} $~MeV, $f_{D_s}/f_D=1.10(2)$, 
$f_{B_s} = 204 \pm 16^{+28}_{-0} $~MeV,
$f_{B} = 179 \pm 18^{+26}_{-9} $~MeV, $f_{B_s}/f_B=1.14(3)^{+0}_{-1}$. Our results, 
which have smaller discretization errors than many previous estimates at fixed value
of the lattice spacing $a$, support a
large value of $f_B$ in the quenched approximation.}
\vfill \eject
\pagestyle{empty}\clearpage
\setcounter{page}{1}
\pagestyle{plain}
\newpage 
\pagestyle{plain} \setcounter{page}{1}
\setcounter{footnote}{0} 
\section{Introduction} \label{sec:intro}
In this paper, we present the results of a lattice calculation
of physical quantities of
phenomenological interest for heavy quarks, such as their mass spectrum 
and decay constants. 
\par In order to reduce the systematic errors, 
we have performed calculations
using the most recent developments in the lattice approach, namely:
\begin{enumerate}
\item \underline{The non-perturbatively improved lattice Clover
action}~\cite{sw}, which we denote as ``Alpha action"~\cite{desy1,desy2,desy3} (see also~\cite{lecture}), 
with the coefficient of the chromomagnetic operator computed in 
Ref.~\cite{desy2};
\item \underline{Non-perturbatively improved vector and axial-vector
currents}, the renormalization coefficients of which have been computed,
using the Ward Identities method~\cite{desy1,wi,clv}, 
in Refs.~\cite{desy2,desy3,sommerguagnelli};
\end{enumerate}
 The use of non-perturbatively improved actions and operators allows us to
reduce the discretization errors to ${\cal O}(a^2)$. This is particularly
important for heavy quark physics since, in current lattice simulations, the typical heavy quark mass $m_Q$ is rather large, $m_Q a \sim 0.3$--$0.6$.
\par Since the coefficient of the Clover term is known non-perturbatively, the hadron spectrum is definitively improved to ${\cal O}(a^2)$.
Unfortunately, the program of removing all the ${\cal O}(a)$ corrections
in the operator matrix elements out of the chiral limit  has not been completed yet, although strategies
to this purpose already exist~\cite{sharperoma,wlee}. 
For this reason, in some cases, we have used  the improvement
coefficients ($b_A$, $c_V$, $b_m$)
evaluated at first order in (boosted) perturbation theory~\cite{sintweisz},
thus leaving us with ${\cal O}(\alpha^2_s\, a m )$ corrections, where $m$
is the relevant quark mass. 
\par
After a careful analysis of the
systematic uncertainties present in the extraction of the physical results,
by assuming quite conservative  errors,  and bearing in mind
 the systematic effects due to the quenched approximation,  the main results of our investigation are the following:  
\begin{itemize}
\item[i)] For $D$ mesons we find:
\bea
f_D = 211 \pm 14 \:{}^{+0}_{-12} \, {\rm MeV},& & 
f_{D_s} = 231 \pm 12 \:{}^{+6}_{-0}  \, {\rm MeV},\quad
{\left( f_{D_s}\over f_{D}\right) } = 1.10(2),  \nn\\  
f_{D^*} = 245 \pm 20 \:{}^{+0}_{-2} \, {\rm MeV},& &
f_{D^*_s} = 272 \pm 16 \:{}^{+0}_{-20}\, {\rm MeV},\quad
{\left( f_{D^*_s}\over f_{D^*}\right) } = 1.11(3) \, ,  
\label{eq:resultd}
\eea
where $f_{D^*}$ and $f_{D_s^*}$ are the vector-meson decay constants.
The latter quantities are not measured experimentally, but enter the calculation of two-body non-leptonic $B$-decays computed using factorization~\cite{neubert}. Thus, they are useful for checking the factorization hypothesis with charmed vector mesons in the final states.  
\item[ii)] For $B$ mesons, we find:
\bea
f_B = 179 \pm 18 \:{}^{+26}_{-9} \, {\rm MeV},& &
f_{B_s} = 204 \pm 16 \:{}^{+28}_{-0}  \, {\rm MeV},\quad
{\left( f_{B_s}\over f_{B}\right) } = 1.14(3)\:{}^{+0}_{-1}, \nn  \\  
f_{B^*} = 196 \pm 24 \:{}^{+31}_{-2} \, {\rm MeV},& &
f_{B^*_s} = 229 \pm 20 \:{}^{+31}_{-16}\, {\rm MeV},\quad
{\left( f_{B^*_s}\over f_{B^*}\right) } =  1.17(4) \:{}^{+0}_{-3}.   \label{eq:fbl} 
\eea
Following Ref.~\cite{bernard}, we have also directly computed the ratio
\beq { f_{B}\over f_{D_s}} = 0.78\pm 0.04\,^{+11}_{-0} \label{eq:fbb}\eeq
from which, using the experimental result,
$f_{D_s}^{(exp.)}= 254 \pm 31\,\mev$~\cite{rich}\footnote{This value has been recently updated by the same authors and reported to us by F.~Parodi.}, we find
\beq 
f_{B}= \frac{f_B}{f_{D_s}} \times f_{D_s}^{(exp.)}=\: (198\pm 24({\rm exp.}) \,^{+30}_{-10}({\rm theo.}) )\; {\rm MeV}\, ,   
\label{aka}
\eeq
Although with a larger error, the result in Eq.~(\ref{aka}) is well compatible with the value given in (\ref{eq:fbl}).
\item[iii)] To reduce the effects of the quenched approximation, we have also used the Grinstein double-ratio method (illustrated below), obtaining
\be r_D= \frac{f_{D_s}}{f_D}= 1.19 (5) \, , \quad  r_B= \frac{f_{B_s}}{f_B}= 1.23 (6) \, , \quad
\frac{f_{B}}{f_{D_s}}= 0.71  \pm 0.04^{+10}_{-0} \ . \label{eq:grin} \ee 
The latter ratio would give as the best estimate for $f_B$:
\be f_B = \: (180\pm 26({\rm exp.}) \,^{+29}_{-10}({\rm theo.}) )\; {\rm MeV}  \,  . \ee 
With the double ratio method we also obtained
\be r_B/r_D=1.03(4) \ .\ee
\item[iv)] We made a detailed study of the hyperfine splitting and of the
scaling laws for masses and
decay constants, as predicted by the heavy quark symmetry. The results of this study can be found below.
\end{itemize}

\noindent
We now give the details of our  analysis and of the methods used to
extract the different physical quantities. Since most of the techniques are
by now standard and have been described {\it ad abundantiam} in the
literature{\footnote{Reviews, with complete lists of references, can be found in~\cite{wittig,sach}.}}, we only focus on those points which are either less common or
new. More details on the calibration of the
lattice spacing and on the extraction of the hadron
masses and matrix elements can be found in Refs.~\cite{spectrum,mescia}.
\par
The remainder of this paper is organized as follows. In Sec.~\ref{sec:details},
we list the main parameters of our simulation and introduce the basic notation
necessary for the discussion of the results. The heavy-light meson
masses and decay constants in lattice units are also given in this  section. 
Since the systematic effects related to the extrapolation/interpolation to the physical point, although related, are quite different in the two cases,
we present separately the physical predictions  for $D$ and $B$ mesons, in
Sec.~\ref{sec:dmesons} and Sec.~\ref{sec:bmesons}, respectively. In Sec.~\ref{sec:scaling}, we discuss the scaling laws predicted by the heavy quark effective theory (HQET) and other related  subjects.
\section{Lattice results}
\label{sec:details}
In this section, we give the essential  information about our
numerical calculation and establish the basic notation.
We then present our results for the heavy-light meson masses and decay constants in lattice units. 
\par The numerical simulation has been performed on a $24^3 \times 64$ lattice,
at $\beta=6.2$, in the quenched approximation. All results and errors
have been obtained with a statistical sample of $100$ independent gauge field
configurations, using the jackknife method with different decimations.
We have used the non-perturbatively improved lattice Clover action,
with $c_{_{SW}}=1.614$~\cite{desy2}.
We work with four values of $\kappa_{light}$, and four $\kappa_{heavy}$:
\begin{itemize}
\item  0.1352 ($\kappa_{\ell_1}$); 0.1349 ($\kappa_{\ell_2}$); 0.1344 ($\kappa_{\ell_3}$); 0.1333 ($\kappa_{\ell_4}$),
\item 0.1250 ($\kappa_{h_1}$); 0.1220 ($\kappa_{h_2}$); 0.1190 ($\kappa_{h_3}$); 0.1150 ($\kappa_{h_4}$). 
\end{itemize}
From the study of the light-hadron spectrum, we obtained
\begin{itemize}
\item[-] $a^{-1}(m_{K^*})=2.75(17)\,\gev$,
\item[-] $\kappa_{crit} = 0.135845(25)$,
\item[-] $\kappa_q = 0.135804(26)$ ,
\item[-] $\kappa_s = 0.13482(12)$ ,
\end{itemize}
where $\kappa_q$, corresponds to the light quark mass $m_q$ (with  $q=u,d$), and $\kappa_s$ to the strange-quark mass,
$m_s$.
The above values have been obtained from the physical pion and kaon masses, by using the method of physical 
lattice planes~\cite{giusti}. All details regarding light hadron spectroscopy and decay constants, can be found in 
Refs.~\cite{spectrum,mescia}.
\par
\begin{figure}
\vspace*{-0.9cm}
\centering
\hspace*{.3cm} \epsfbox{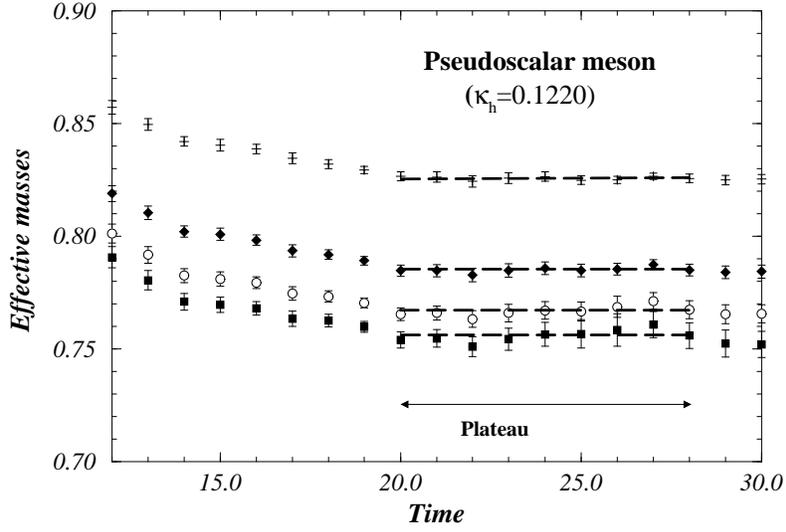}
\vspace{0.5cm}
\hspace*{.3cm} \epsfbox{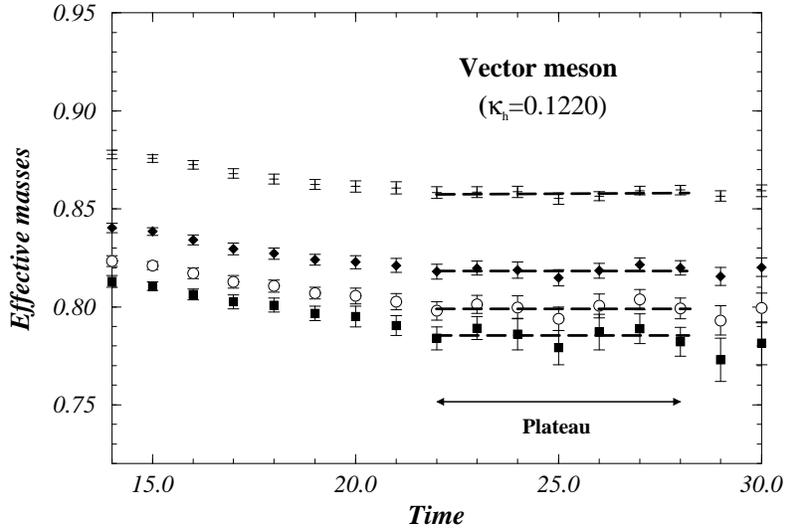}
\caption{\it Effective masses of heavy-light pseudoscalar and vector mesons as a function of the time in lattice units. In each figure, the heavy quark mass (corresponding to $\kappa_h=0.1220$) is fixed, and combined with four different light quark masses.}
\label{fig:plateau}
\end{figure}

 For the mass spectrum, following the standard procedure,
we measured suitable  two-point correlation functions, from which we can 
isolate the lowest lying states
\bea
C_{_{JJ}}(t) = \sum_{\vec x} \langle 0 \vert J({\vec x}, t)
 J^{\dagger}(0)\vert 0 \rangle\, \stackrel{t\gg 0}{\longrightarrow}\, {{\cal{Z}}_J \over  {\rm sinh}(M_J)}e^{- M_J T/2}
  {\rm cosh}\left[ M_J \left( {T\over 2} - t \right) \right]  ,
\label{eq:meff}
\eea
where $J\equiv J_{PS}=\bar Q \gamma_5 q$, or $J\equiv J^k_V=\bar Q \gamma^k q$.
In Fig.~\ref{fig:plateau}, we show the effective masses for the pseudoscalar and vector heavy-light mesons
at fixed heavy quark mass. By inspection, we established the fit intervals 
$t\in [20,28]$,  and $t\in [22,28]$, for the pseudoscalar and vector cases, respectively.
The resulting pseudoscalar and vector masses in lattice units, as well as the matrix elements,
${\cal Z}_{PS}=\vert \langle  PS(\vec p=0) \vert J_{PS} \vert 0 \rangle \vert^2$ and 
${\cal Z}_{V}=\vert \langle  V(\vec p=0;\lambda) \vert J^k_{V} \vert 0 \rangle \vert^2$, 
are listed in Tab.~\ref{tab:masses}.
\begin{table}[h]
\centering
\begin{tabular}{|c|c|c|c|c|} \hline
{\phantom{\huge{l}}}\raisebox{-.2cm}{\phantom{\Huge{j}}}
{  ``Flavor" content}& { ${M_{PS}}$} & { ${\cal{Z}}_{ PS}$} & ${ M_{V}}$ & { ${\cal{Z}}_{V}$} \\ \hline \hline
{\phantom{\Large{l}}}\raisebox{.2cm}{\phantom{\Large{j}}}
{\hspace{-1.5mm}$h_4-\ell_4$\hspace{1.5mm}} & 1.0256(19) & 0.0254(9) & 1.0489(21) & 0.0120(6) \\
{ $h_4-\ell_3$} & 0.9868(29) & 0.0198(11) & 1.0104(33) & 0.0091(7) \\
{ $h_4-\ell_2$} & 0.9696(45) & 0.0176(16) & 0.9920(52) & 0.0077(9) \\
{\phantom{\Large{l}}}\raisebox{-.2cm}{\phantom{\Large{j}}}
{\hspace{-1.5mm}$h_4-\ell_1$\hspace{1.5mm}} & 0.9584(67) & 0.0158(22) & 0.9783(77) & 0.0065(11) \\ \hline
\hline
{\phantom{\Large{l}}}\raisebox{.2cm}{\phantom{\Large{j}}}
{\hspace{-1.5mm}$h_3-\ell_4$\hspace{1.5mm}}  & 0.9143(17) & 0.0237(8) & 0.9420(20) & 0.0105(5) \\
{ $h_3-\ell_3$} & 0.8746(25) & 0.0186(9) & 0.9032(32) & 0.0080(6) \\
{ $h_3-\ell_2$} & 0.8569(38) & 0.0166(13) & 0.8844(49) & 0.0068(7) \\
{\phantom{\Large{l}}}\raisebox{-.2cm}{\phantom{\Large{j}}}
{\hspace{-1.5mm}$h_3-\ell_1$\hspace{1.5mm}}  & 0.8458(56) & 0.0150(18) & 0.8705(72) & 0.0057(9) \\ \hline
  \hline
{\phantom{\Large{l}}}\raisebox{.2cm}{\phantom{\Large{j}}}
{\hspace{-1.5mm}$h_2-\ell_4$\hspace{1.5mm}} & 0.8256(16) & 0.0221(7) & 0.8577(20) & 0.0093(4) \\
{ $h_2-\ell_3$} & 0.7851(23) & 0.0175(8) & 0.8185(31) & 0.0071(5)  \\
{ $h_2-\ell_2$} & 0.7669(34) & 0.0157(11) & 0.7994(47) & 0.0060(6) \\
{\phantom{\Large{l}}}\raisebox{-.2cm}{\phantom{\Large{j}}}
{\hspace{-1.5mm}$h_2-\ell_1$\hspace{1.5mm}} & 0.7558(48) & 0.0144(15) & 0.7853(68) & 0.0051(8) \\ \hline
\hline
{\phantom{\Large{l}}}\raisebox{.2cm}{\phantom{\Large{j}}}
{\hspace{-1.5mm}$h_1-\ell_4$\hspace{1.5mm}} & 0.7304(13) & 0.0199(5) & 0.7683(21) & 0.0079(3) \\
{ $h_1-\ell_3$} & 0.6894(19) & 0.0161(6) & 0.7295(30) & 0.0062(3) \\
{ $h_1-\ell_2$} & 0.6707(27) & 0.0145(9) & 0.7099(42) & 0.0052(4) \\
{\phantom{\Large{l}}}\raisebox{-.2cm}{\phantom{\Large{j}}}
{\hspace{-1.5mm}$h_1-\ell_1$\hspace{1.5mm}} & 0.6594(37) & 0.0136(12) & 0.6960(57) & 0.0045(5) \\
\hline \end{tabular}
\vspace*{1.2cm}
\caption{\it Mass spectrum of heavy-light pseudoscalar and vector mesons in lattice units.}
\label{tab:masses}
\end{table}
\newpage
\par We used the standard  procedure to extract the pseudoscalar and vector decay constants. This procedure consists in calculating the ratios
\bea
{{\sum_{\vec x} \langle \hat{A}_0(\vec x,t) P(0) \rangle } 
\over {\sum_{\vec x} \langle P(\vec x,t) P(0) \rangle }}  &\simeq& 
 {\hat{F}}_P \, {M_P\over {\sqrt {\cal{Z}}_P}}\, {\rm  tanh} \left(M_P (\frac{T}{2} - t)\right) \\
\vspace*{-.6cm} {\sum_{\vec x} \langle \hat{V}_i(\vec x,t) \hat{V}_i(0)\rangle } 
 &\simeq&  M_V^2 {\hat{F}}_V^2 e^{-M_V {T \over 2}}\, {\rm  cosh} \left(M_V (\frac{T}{2} - t)\right)
\eea
where we assumed the usual definitions
\bea
\langle 0|{\hat{A}}_0|PS(\vec p = 0)\rangle &=& i {\hat{F}}_{PS} M_{PS},\hspace*{6mm} {\rm{and}}\nonumber \\
\langle 0|{\hat{V}}_i|V(\vec p = 0;\lambda )\rangle &=& i e_i^{(\lambda)} {\hat{F}}_V M_V.
\eea
We denote  decay constants and meson masses in lattice units  by capital letters,
and the hat reminds us that the quantity is improved and {\it renormalized}. 
In practice, one first partially 
improves the bare lattice currents (for clarity, we write the lattice spacing $a$  explicitly):
\bea
\langle 0|A_0|PS(\vec p = 0)\rangle &\to& \langle 0|A_0|PS\rangle + c_A \langle 0|a \partial_0 P|PS\rangle 
= i M_{PS} (F_{PS}^{(0)} + c_A a F_{PS}^{(1)}),\cr
\hfill \cr
\langle 0|V_i|V(\vec p = 0;\lambda )\rangle &\to& \langle   0|V_i|V\rangle + c_V  \langle 0|a \partial_0 T_{i0}|V\rangle = i M_V e_i^{(\lambda)} (F_V^{(0)}+ c_V a F_V^{(1)}),
\eea
and then multiplies the currents by suitable overall factors  
\bea
{\hat F}_P &=& Z_A \, ( 1 + b_A a m )\,  (F_P^{(0)} \, + \, c_A a F_P^{(1)}),\quad \left[Z_A(m)=
Z_A \, ( 1 + b_A a m )\right]\cr
{\hat F}_V &=& Z_V \, ( 1 + b_V a m )\,  (F_V^{(0)} \, + \, c_V a F_V^{(1)}), \quad \left[Z_V(m)=
Z_A \, ( 1 + b_V a m )\right].
\label{eq:rco}
\eea
In the calculation of the different correlations above, when the lowest state is well isolated, we may use
\bea
{\langle \partial_0 P(t) P(0)\rangle \over \langle P(t) P(0)\rangle}\, = \, {\rm  sinh}( M_{PS} ),
\eea
\beq
a F_{PS}^{(1)} =  {{\sqrt {{\cal{Z}}_{PS}}}\over M_{PS}}\,  {\rm  sinh} (M_{PS})\, , 
\hspace*{6mm} {\rm{and}}\hspace*{6mm}
a F_V^{(1)} =  {{\sqrt {{\cal{Z}}_{T_i}}}\over M_V}\,  {\rm  sinh}(M_V).
\eeq
The values of the decay constants are given in Tab.~\ref{tab:dco}.\\
\begin{table}[h]
\centering
\begin{tabular}{|c|ccc|ccc|} \hline 
{\phantom{\huge{l}}}\raisebox{-.3cm}{\phantom{\Huge{j}}}
{\hspace*{-4.5mm} $\kappa_1\,\kappa_2$}& { $F_{PS}^{(0)}$} & { $- c_{_A} a F_{PS}^{(1)}/F_{PS}^{(0)}$} &  { $F_{PS}$} & {  $F_{V}^{(0)}$} & { $- c_{_V} a F_{V}^{(1)}/F_{V}^{(0)}$} &  { $F_{V}$} \\ \hline \hline
{\phantom{\Large{l}}}\raisebox{.3cm}{\phantom{\Large{j}}}
{\hspace*{-1.5mm}$h_4$-$\ell_4$\hspace{1.5mm}} & 0.0957(16) & 0.0730(6) &{\sf 0.0887(15)} & 0.1043(21) & 0.0276(2) &{\sf 0.1014(21)}\\

{\phantom{\Large{l}}}\raisebox{-.05cm}{\phantom{\Large{j}}}
{ \hspace{-4mm}$h_4$-$\ell_3$} & 0.0869(23) & 0.0702(6) & {\sf 0.0917(29)} & 0.0942(30) & 0.0260(4) &{\sf  0.0740(23)} \\
{\phantom{\Large{l}}}\raisebox{-.1cm}{\phantom{\Large{j}}}
{ \hspace{-4mm}$h_4$-$\ell_2$} & 0.0825(35) & 0.0694(8) &{\sf  0.0859(42)}& 0.0881(44) & 0.0251(6) &{\sf  0.0698(34)} \\
{\phantom{\Large{l}}}\raisebox{-.3cm}{\phantom{\Large{j}}}
{\hspace{-4mm} $h_4$-$\ell_1$} & 0.0793(48) & 0.0687(10) & {\sf 0.0803(58)} & 0.0823(60) & 0.0245(9) &{\sf 0.0658(48)} \\ 
\hline

{\phantom{\Large{l}}}\raisebox{.3cm}{\phantom{\Large{j}}}
{\hspace*{-1.5mm}$h_3$-$\ell_4$\hspace{1.5mm}}& 0.0982(15) & 0.0664(6) & {\sf 0.0917(14)} & 0.1089(21) & 0.0235(2) & {\sf 0.1063(20)}\\
{\phantom{\Large{l}}}\raisebox{-.05cm}{\phantom{\Large{j}}}
{\hspace{-3mm} $h_3$-$\ell_3$} & 0.0896(21) & 0.0639(6) &{\sf 0.0839(20)} & 0.0990(30) & 0.0221(3) & {\sf 0.0969(29)}  \\
{\phantom{\Large{l}}}\raisebox{-.1cm}{\phantom{\Large{j}}}
{ \hspace{-3mm}$h_3$-$\ell_2$} & 0.0853(30) & 0.0630(8) &{\sf 0.0799(28)}& 0.0928(42) & 0.0214(5) &{\sf 0.0908(41)}  \\
{\phantom{\Large{l}}}\raisebox{-.3cm}{\phantom{\Large{j}}}
{\hspace{-4mm} $h_3$-$\ell_1$}  & 0.0822(42) & 0.0625(10) & {\sf 0.0771(39)} & 0.0870(58) & 0.0208(7) & {\sf 0.0852(57)} \\ \hline

{\phantom{\Large{l}}}\raisebox{.3cm}{\phantom{\Large{j}}}
{\hspace*{-1.5mm}$h_2$-$\ell_4$\hspace{1.5mm}}& 0.0999(14) & 0.0615(6) & {\sf 0.0938(14)} & 0.1126(21) & 0.0206(2) & {\sf 0.1103(20)}\\
{\phantom{\Large{l}}}\raisebox{-.05cm}{\phantom{\Large{j}}}
{\hspace{-3.85mm} $h_2$-$\ell_3$} & 0.0916(19) & 0.0591(6) & {\sf 0.0862(18)} & 0.1032(29) & 0.0193(3) & {\sf 0.1012(29)}  \\
{\phantom{\Large{l}}}\raisebox{-.1cm}{\phantom{\Large{j}}}
{\hspace{-.5mm} $h_2$-$\ell_2$}\hspace{3.mm} & 0.0875(27) & 0.0583(8) &{\sf  0.0824(25)}& 0.0969(41) & 0.0186(4) &{\sf  0.0951(40)}  \\
{\phantom{\Large{l}}}\raisebox{-.3cm}{\phantom{\Large{j}}}
{\hspace{-4mm} $h_2$-$\ell_1$} & 0.0846(37) & 0.0578(9) & {\sf 0.0797(35) }& 0.0910(57) & 0.0181(6) & {\sf 0.0894(56) }\\ \hline

{\phantom{\Large{l}}}\raisebox{.3cm}{\phantom{\Large{j}}}
{\hspace*{-1.5mm}$h_1$-$\ell_4$\hspace{1.5mm}}& 0.1009(15) & 0.0564(5) & {\sf 0.0953(15)} & 0.1154(20) & 0.0178(1) & {\sf 0.1134(19)}\\
{\phantom{\Large{l}}}\raisebox{-.05cm}{\phantom{\Large{j}}}
{ \hspace{-3mm}$h_1$-$\ell_3$} & 0.0933(18) &  0.0544(6) & {\sf 0.0883(18)} & 0.1078(29) & 0.0165(2) & {\sf 0.1060(29)}  \\
{\phantom{\Large{l}}}\raisebox{-.1cm}{\phantom{\Large{j}}}
{ \hspace{-3mm}$h_1$-$\ell_2$} & 0.0895(24) & 0.0536(8) &{\sf  0.0847(23)}& 0.1015(40) & 0.0159(3) &{\sf  0.0999(39)} \\
{\phantom{\Large{l}}}\raisebox{-.3cm}{\phantom{\Large{j}}}
{\hspace{-4mm} $h_1$-$\ell_1$} & 0.0870(32) & 0.0531(9) & {\sf 0.0823(30)} & 0.0957(55) & 0.0155(5) & {\sf 0.0942(54)} \\ \hline

\end{tabular}
\caption{{\it Heavy-light decay constants in lattice units.}}
\label{tab:dco}
\end{table}
The improvement coefficients and the renormalization constants 
are catalogued in {Tab.~3,} where we  
\begin{table}[h]
\centering
\begin{tabular}{|c||c|c|} \hline 
\multicolumn{3}{|c|}
{\raisebox{.3cm}
{\phantom{\huge{l}}}\raisebox{-.2cm}{\phantom{\Huge{j}}}
$\stackrel{\mbox{{\sf Renormalization constants}}}{\mbox{{\sf (in the chiral limit)}}}
$}
\\  \hline
{\sl Quantity} & $Z_V$ & $Z_A$ \\
{\sl BPT} & 0.846 & 0.862 \\
${\sl Non-perturbative}$ & {\bf 0.793} &{\bf 0.809} \\ \hline \hline
\multicolumn{3}{|c|}
{\raisebox{.3cm}
{\phantom{\Huge{l}}}\raisebox{-.2cm}{\phantom{\Huge{j}}}
$\stackrel{\mbox{{\sf Coefficients for the Improvement}}}{\mbox{{\sf  of the Bare Operators}}}
$}
\\  \hline
{\sl Quantity}  & $c_V$ & $c_A$ \\
{\sl BPT} & {\bf -0.026} & -0.012\\
{\sl Non-perturbative} & -0.214(74) &{\bf -0.037}  \\ \hline \hline
\multicolumn{3}{|c|}
{\raisebox{.3cm}
{\phantom{\Huge{l}}}\raisebox{-.2cm}{\phantom{\Huge{j}}}
$\stackrel{\mbox{{\sf Coefficients for the Renormalization Constants}}}{\mbox{{\sf Improvement (due to the Explicit Mass Term) }}}
$}
\\  \hline
{\sl Quantity}  & $b_V$ & $b_A$ \\
{\sl BPT} & 1.242 & {\bf 1.240} \\
{\sl Non-perturbative} &{\bf 1.404} &{\sc not calc.}  \\ \hline 
\end{tabular}
{\caption{\it Improvement coefficients. In boosted perturbation theory $g^2=1.256$.
 For the perturbative $Z_J$'s, we used $c_{_{SW}}=1.614$. The values which have been used in our numerical calculations are marked in bold.}}
\label{tab:lista}
\end{table}
also display the one-loop results obtained by using 
boosted perturbation theory (BPT) at $\beta=6.2$~\cite{sintweisz,capitani}\footnote{This corresponds to the use of $g^2=1.256$, in the perturbative formulae.}.  
Recall that the corrective coefficients $b_J$ 
enter with the ``average" quark mass defined as $am = am_{ij} = {1\over 2} \left( am_i + am_j \right)$, where the bare mass is the one derived from the vector Ward identity, namely 
\be am_i = {1\over 2} \left( {1 \over \kappa_i} -{1\over \kappa_{crit}}\right) \ . \label{eq:qmasses}
\ee
In the following, we denote by $m_q$ and
$m_Q$ the generic light and heavy quark masses, whereas the quark masses expressed in terms of the corresponding hopping parameters, as in Eq.~(\ref{eq:qmasses}), are denoted by $m_\ell$ or $m_h$.
\par 
Note that, in spite of the non-perturbative determination of $c_{V}$, we used the perturbative value $c_{V}^{\scr{BPT}} = -0.026$. Firstly, we find the
non-perturbative result  of Ref.~\cite{sommerguagnelli}, $c_{V}^{\scr{NP}} = -0.214(74)$, surprising because it is one order of magnitude larger than $c_{V}^{\scr{BPT}}$.
This possibility is not excluded {\it a priori}, but it is difficult to accommodate it in the pattern of all other improvement coefficients: when known non-perturbatively, their value is always close to the  corresponding (boosted) perturbative one and never differs by one order of magnitude. Secondly, by using $c_{V}^{\scr{NP}} = -0.214(74)$, the ratio  of the vector to the pseudoscalar meson decay constants $f_{H^*}/f_H$ badly fails in approaching {\it one}, as $M_H$ increases, contrary to what is predicted by heavy quark symmetry. More details on this scaling law will be given in Sec.~\ref{sec:scaling}. For these reasons, we find it safer to use the $c_{V}^{\scr{BPT}}$. We believe that the preliminary determination of $c_{V}^{\scr{NP}}$ in Ref.~\cite{sommerguagnelli} has some problem and prefer to wait for the final results.

\section{$D$-meson spectrum and decay constants}
\label{sec:dmesons}
In this section, we discuss the $D$-meson spectrum
and  decay constants. 
Preliminary results of this study were given in Ref.~\cite{boulder}\footnote{See~\cite{ukqcdboulder} for preliminary results from the UKQCD collaboration and the APETOV group.}.
\par
\begin{figure}[h]
\centering
\hspace*{0.3cm}\epsfbox{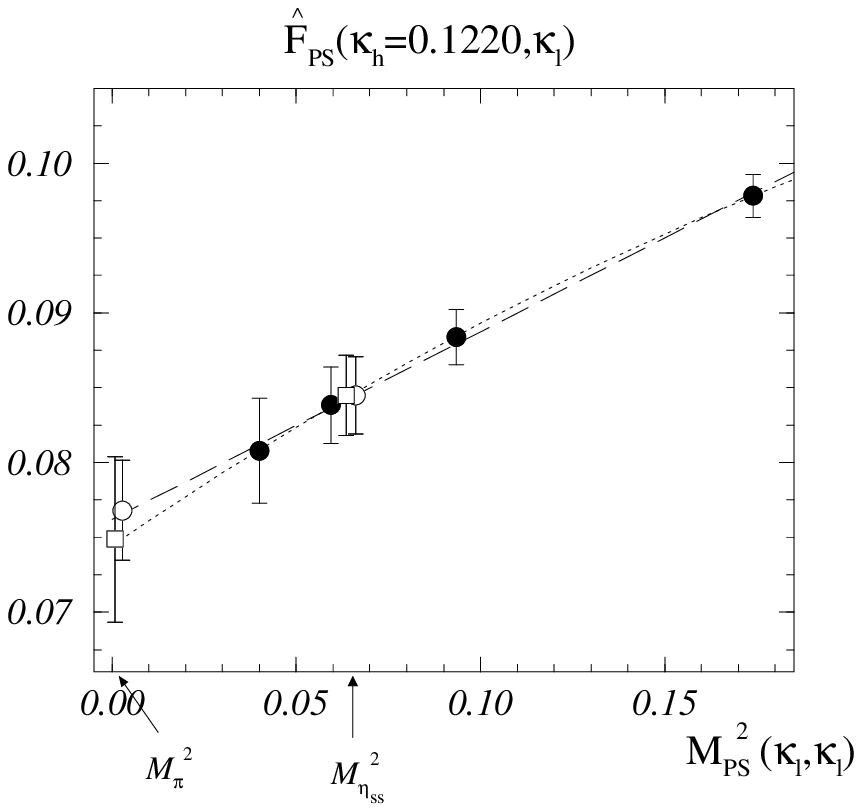}
\caption{{\it Fits of $\hat{F}_{PS}$ in the light quark mass,  at fixed $m_h$. The filled circles
denote data directly measured.
The  dashed curve and  empty circles refer to the linear fit and extrapolated points. The  
dotted curve and empty squares to the quadratic fit and extrapolated points. 
The heavy quark mass corresponds  to $\kappa_{h_2}$.}}
\label{fig:lightextr}
\end{figure}

 
In Tab.~\ref{tab:lfit}, we tabulate the results for the heavy-light meson masses, $M_H(m_h,m_\ell)$,
obtained from a linear extrapolation (interpolation) in the light quark mass (to reach $q=u,d$  and $s$).
\begin{table}[h]
\centering
\begin{tabular}{|c|c|c|c|c|} 
\multicolumn{5}{c}
{\raisebox{-.1cm}
{\phantom{\huge{l}}}\raisebox{-.2cm}{\phantom{\Huge{j}}}
}
\\  \hline
{\phantom{\huge{l}}}\raisebox{-.2cm}{\phantom{\Huge{j}}}
{  ``Flavor" content}& { ${\rm M_{PS}}$} & { $\hat{F}_{\rm PS}$} & ${\rm M_{V}}$ & { $\hat{F}_{\rm V}$} \\ \hline \hline
{\phantom{\Large{l}}}\raisebox{.2cm}{\phantom{\Large{j}}}
{\phantom{\Large{l}}}\raisebox{+.2cm}{\phantom{\Large{j}}}
{ \hspace{-5mm}$h_4$--$s$\hspace{1mm}} & {\sf 0.9721(61)} & {\sf 0.0870(36)} & {\sf 0.9947(62)} & {\sf 0.0992(49)} \\    
{\phantom{\Large{l}}}\raisebox{+.2cm}{\phantom{\Large{j}}}
{$h_4$--$q$} & {\sf 0.9414(58)} & {\sf 0.0783(47)} & {\sf 0.9635(65)} & {\sf 0.0875(61)} \\ \hline  
{\phantom{\Large{l}}}\raisebox{.2cm}{\phantom{\Large{j}}}
{\phantom{\Large{l}}}\raisebox{+.2cm}{\phantom{\Large{j}}}
{ \hspace{-5mm}$h_3$--$s$\hspace{1mm}} & {\sf 0.8595(56)} & {\sf 0.0856(30)} & {\sf 0.8871(61)} & {\sf 0.0986(45)} \\    
{\phantom{\Large{l}}}\raisebox{.2cm}{\phantom{\Large{j}}}
{$h_3$--$q$}  & {\sf 0.8281(50)} & {\sf 0.0774(39)} & {\sf 0.8555(62)} & {\sf 0.0877(56)} \\ \hline   
{\phantom{\Large{l}}}\raisebox{.2cm}{\phantom{\Large{j}}}
{\phantom{\Large{l}}}\raisebox{+.2cm}{\phantom{\Large{j}}}
{ \hspace{-5mm}$h_2$--$s$\hspace{1mm}}  &{\sf  0.7697(53)} & {\sf 0.0845(26)} & {\sf 0.8022(61)} & {\sf 0.0985(43)} \\   
{\phantom{\Large{l}}}\raisebox{.2cm}{\phantom{\Large{j}}}
$h_2$--$q$  & {\sf 0.7375(43)} & {\sf 0.0768(33)} & {\sf 0.7702(59)} & {\sf 0.0881(53)} \\ \hline  
{\phantom{\Large{l}}}\raisebox{.2cm}{\phantom{\Large{j}}}
{ \hspace{1mm}$h_1$--$s$\hspace{1mm}}  &{\sf 0.6736(49)} & {\sf 0.0831(22)} & {\sf 0.7126(62)} & {\sf 0.0985(40)} \\ 
{\phantom{\Large{l}}}\raisebox{.2cm}{\phantom{\Large{j}}}
{$h_1$--$q$}  & {\sf 0.6409(35)} & {\sf 0.0761(29)} & {\sf 0.6804(54)} & {\sf 0.0896(51)} \\ \hline  
\end{tabular}
\vspace*{.8cm}
\caption{{\it Mass spectrum and decay constants of heavy-light pseudoscalar and vector mesons. $h_i$--$s$ 
and $h_i$--$q$ denote mesons composed
by a heavy quark with mass $m_{h_i}$ and a strange or a light ($u,d$) quark, respectively. All the results are expressed in lattice units.}}
\label{tab:lfit}
\end{table}
This was achieved by using the method of physical lattice planes. In Ref.~\cite{spectrum}, we extracted the $M_\pi$, and the hypothetical pseudoscalar $M_{\eta_{ss}}$, which (when squared) are proportional to $m_q$ and $m_s$, respectively. For the generic physical quantity in the heavy-light meson sector~${\frak{F}}_H(m_h,m_\ell)$, we use the following form of fit  
\bea
{\frak{F}}_{H}(m_h,m_{\ell_i}) = \alpha_h + \beta_h M_{PS}^2(m_{\ell_i},m_{\ell_i})
+ \gamma_h \left( M_{PS}^2(m_{\ell_i},m_{\ell_i})\right)^2,
\label{eq:impr_lin}
\eea
where the heavy quark mass ({\it i.e.} $\kappa_h$) is kept fixed. The coefficients of such a fit, $\alpha_h$, $\beta_h$, $\gamma_h$, are then used to obtain ${\frak{F}}_{H}(m_h,m_{q})$ and ${\frak{F}}_{H}(m_h,m_{s})$, by inserting on the {\it r.h.s.} of~(\ref{eq:impr_lin}), $M_{\pi}^2$ and $M_{\eta_{ss}}^2$, respectively.
In practice, it turns out that the linear ($\gamma_h = 0$) and quadratic ($\gamma_h \neq 0$) fits give essentially the same results for any
physical quantity considered in this study~\footnote{As expected, the results 
obtained from a quadratic fit inflates the errors in extrapolated results.}. 
In Fig.~\ref{fig:lightextr}, we show this effect for the pseudoscalar decay constant.
Therefore, in all results that we present in what follows, whenever a quantity with light quark flavor $q$ and/or $s$ is mentioned, it means that the linear fit in Eq.~(\ref{eq:impr_lin}) is performed, {\it i.e.} $\gamma_h=0$.

Having fixed the light quark mass, we now want to interpolate in the heavy quark mass.
In the framework of the HQET, the functional dependence of $M_H$ on $m_Q$ is: 
\bea
M_H = m_Q \left\{ 1 + {{\bar{\Lambda}}\over m_Q} + {1 \over 2 m_Q^2} 
\left( \lambda_1 + k \lambda_2 \right) + {\cal{O}}({1\over m_Q^3})\right\}
\label{eq:rel5}
\eea
where ${\bar{\Lambda}}$ is the so-called binding energy, 
$\lambda_{1,2}$ are the first (flavor-spin) symmetry breaking corrections 
(describing the kinetic and chromomagnetic energy), and $k=3(-1)$, for {${J^P=0^-(1^-)}$.} 
The improvement of the quark mass brings in the quadratic 
terms in $m_h$, {\it i.e.} $m_h \to m_h ( 1 + b_m m_h)$, and distort all the coefficients in the expansion (\ref{eq:rel5}). The term of order $m_h^2$   originates only from the lattice artifacts, and thus is always proportional to $b_m$. The interplay between power corrections in $1/m_Q$ and discretization effects, however, modifies the ``effective" value of $b_m$, {\it i.e.} the coefficient of the quadratic term in $m_h$. 
To investigate this point, we study the behaviour of $M_H$ in $m_h$, at fixed light quark mass, $m_q$. In the $D$ case, we use the following expression:
\begin{eqnarray}
 M_H (m_{h_i},m_q) - {a\,m_D} = {\cal{A}} \left( {m_{h_i}} 
 - {m_{charm}} \right) \left[ 1 + {\cal{B}} \left( {m_{h_i}} 
 + {m_{charm}} \right)
\right]
\label{eq:fform}
\end{eqnarray}
where   $a\,m_{D}$ is the experimental meson mass  in lattice units, $a m_D= 0.68(4)$
(similarly we fit $M_{H^*} (m_{h_i},m_q) - {a\,m_{D^*}}$, etc.).
From the fit of our data to Eq.~(\ref{eq:fform}),
it turns out that the resulting value for $\kappa_{charm}$ is 
stable for ${\cal{B}}\in [-0.4,-0.2]$. The minimum $\chi^2$ is
reached for ${\cal{B}}=-0.32$. We have also performed the linear fit (corresponding to ${\cal{B}}=0$),
and the fit with ${\cal{B}}=b^{^{BPT}}_m=-0.652$~\cite{sintweisz}. The different values that we obtain for $\kappa_{charm}$ 
with different fits (linear, quadratic or using $b^{^{BPT}}_m$) differ 
by about one per mille. We quote
\bea
\kappa_{charm}= 0.1231(14) .
\label{eq:kch}
\eea
It can be argued that a fit of  the spin-average mass ${\overline{M}}_H=(3 M_V + M_{PS})/4$, 
to extract $\kappa_{charm}$ is more suitable, because  spin forces of ${\cal O}(1/m_Q)$  are canceled 
in this combination (see~(\ref{eq:rel5})).
For ${\cal{B}}=-0.32$, corresponding also in this case to the 
minimum $\chi^2$, 
we obtain $\kappa_{charm}=0.1232(14)$. Since the differences for the $D-$meson masses and decay constants as obtained by using the two values of $\kappa_{charm}$ is very small, in the following, whenever we refer to $\kappa_{charm}$, the  value~(\ref{eq:kch}) is understood. 
Using $M_D$ as a physical input (to fix $\kappa_{charm}$), we can make several  predictions for other meson masses
\bea
M_{D}\equiv{\rm input} &;& M_{D^*}=0.725(42),\quad {\rm and}\nonumber \\
M_{D_s}=0.733(46)&;&  M_{D_s^*}=0.768(45).
\eea
\noindent
which in physical units give
\be m_{D^*}= 1.992(24)\,\gev\, , \quad m_{D_s}=2.013(18)\,\gev\, , \quad m_{D_s^*}=2.110(21)\,\gev \ , \ee
to be compared to the experimental numbers~\cite{experiment}
\be m^{(exp.)}_{D^*}=2.008\,\gev;\quad m^{(exp.)}_{D_s}=1.968\,\gev;\quad m^{(exp.)}_{D_s^*}=2.112(27)\,\gev \ .\ee 
We obviously fail to obtain the experimentally measured mass-difference. We get  \bea
m_{D_s^*}-m_{D_s} = (97\pm 12)\,\mev.
\eea
which is to be compared to $(m_{D_s^*}-m_{D_s})^{(exp.)} = 143.8(4)\,\mev$.
\par
\item An alternative procedure 
is to  consider the ratio  $C_{_{VV}}/C_{_{PP}}$, from which  the vector-pseudoscalar mass difference  can be directly
extracted. By using this method we get
\bea
M_{D^*}-M_{D}=0.0354(39) &\to& m_{D^*}-m_{D}=97(15)\, \mev,\nonumber \\
\hfill \nonumber \\
M_{D_s^*}-M_{D_s}=0.0354(30) &\to& m_{D_s^*}-m_{D_s} = 97(13) \, \mev ;\nonumber \eea
which confirms that the ({\sl `spin'}) mass difference is {\bf systematically} smaller than
the experimental one, regardless of the procedure we use. 
Since we found a reasonable agreement for the hyperfine splitting in the light-quark sector~\cite{spectrum}, 
the discrepancy in the heavy-quark case  is probably a signal of large residual  ${\cal{O}}(a^2)$ errors.
We believe that the discrepancy cannot be entirely attributed to the use of the quenched approximation~\footnote{Quenching is always a joker-argument when we are unable to solve or explain a problem in lattice calculations.}.
\par
\begin{figure}[h]
\centering
\hspace*{0.3cm}\epsfbox{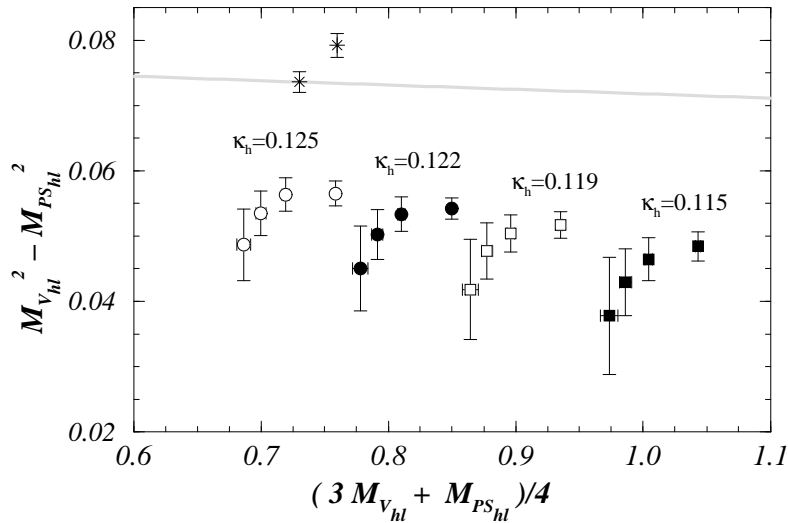}
\caption{{\it Hyperfine splitting for heavy-light mesons in 
lattice units. The gray line shows the (approximate) experimental slope.
The stars mark the physical $m_{D^*}^2 - m_{D}^2$  and  $m_{D_s^*}^2 - m_{D_s}^2$ splittings.}}
\label{fig:split1}
\end{figure}

 Our results for the hyperfine splitting are shown in Fig.~\ref{fig:split1}. From that figure, we note the qualitative
agreement between the dependence of the splitting on the light-quark mass measured on the lattice and its experimental counterpart. Moreover, the  dependence of the hyperfine splitting on the meson mass is not dramatically larger than the experimental one, represented by a gray line in the figure. This is to be contrasted to the case of the unimproved Wilson action, where the lattice slope is by far larger than in the present case~\cite{bmasc}, showing a clear effect of  improvement, although insufficient to describe the experimental numbers. In Tab.~\ref{tab:tabla9}, we list the results extrapolated in the light quark mass, at fixed $m_h$.
\begin{table}[h]
\centering
\begin{tabular}{|c|c|c|} \hline 
{\phantom{\Huge{l}}}\raisebox{-.2cm}{\phantom{\Huge{j}}}
{ heavy} & { $q$}& { $s$} \\ \hline 
{\phantom{\Huge{l}}}\raisebox{-.1cm}{\phantom{\Huge{j}}}
{$h_1$} & 0.393(53) & 0.406(50)   \\
{\phantom{\Huge{l}}}\raisebox{-.1cm}{\phantom{\Huge{j}}}
{$h_2$} & 0.367(57) & 0.383(52)  \\ 
{\phantom{\Huge{l}}}\raisebox{-.1cm}{\phantom{\Huge{j}}}
{$h_3$} & 0.342(61) & 0.361(54)   \\
{\phantom{\Huge{l}}}\raisebox{-.1cm}{\phantom{\Huge{j}}}
{$h_4$} & 0.310(68) & 0.330(57)  \\ \hline
\end{tabular}
\caption{{\it Hyperfine splitting, ${M_{H^*}^2 - M_H^2}$ in $\gev^2$.}}
\label{tab:tabla9}
\end{table}
\par We now discuss the physical results for the $D$-meson decay constants. 
We first extrapolate the results, in the light quark mass, at fixed $m_h$, by using 
 a formula similar to Eq.~(\ref{eq:impr_lin}), {\it i.e.} 
\bea
\hat  F_{H}(m_h,m_{\ell_i}) = \alpha_h^{\scr F}\: +\: \beta_h^{\scr F}\, M_{PS}^2(m_{\ell_i},m_{\ell_i})\ .
\label{eq:impr_linf}
\eea
The results of this extrapolation, $m_{\ell_i}\to m_{q}$ and $m_{\ell_i}\to m_{s}$, are given in Tab.~\ref{tab:lfit}.
\par To handle the problem of extrapolation in the heavy-quark mass, at fixed light-quark mass,
we rely on the heavy quark symmetry. The relevant scaling law is
\bea
f_H =  {\Phi(m_H) \over \sqrt{m_H}} \left( 1 + \frac{\Phi^\prime(m_H)}{\Phi(m_H)\, m_H} + \dots \right) \  ,
\eea
where $\Phi(m_H)$, $\Phi^\prime(m_H)$ depend logarithmically on the mass, e.g. $\Phi(m_H) \sim
\alpha_s^{-2/b_0}(m_H) (1 + {\cal{O}}(\alpha_s ) )$~\footnote{ We prefer to give the scaling law in terms of the hadron mass $m_H$ rather than the heavy quark mass.}. In the interval of masses considered in this study, the logarithmic corrections are negligible. For this reason, in our fits, we used 
\bea
{\hat F}_H \sqrt{M_H} =  \Phi_0 + {\Phi_1 \over M_H}+ {\Phi_2 \over M_H^2}\ ,
\label{eq:scaling}
\eea
where $\Phi_0$, $\Phi_1$ and $\Phi_2$ are constants.   
At the physical point $M_H= a m_D$ (corresponding to $\kappa_h=\kappa_{charm}$), we read off the value $\hat F_D$ in lattice units. To express it in physical units, one simply multiplies by $a^{-1}$. The same procedure
can be used for the vector-meson decay constants. 
\par Another possibility,
is to consider the ratios $\hat R_H(m_h,m_{\ell})= \hat F_H(m_h,m_{\ell})/\hat F_{PS}(m_q,m_{\ell})$ and $\hat R_{H^*}= \hat F_{H^*}(m_h,m_{\ell})/\hat F_V(m_q, m_{\ell})$, and to extrapolate $\hat R_H(m_h,m_\ell)$ in $m_\ell$ and $m_h$ by using eqs.~(\ref{eq:impr_linf}) and (\ref{eq:scaling}), with the obvious replacement
$\hat F_H \to \hat R_H$ ($\hat F_{\pi,\rho}=\hat F_{PS,V}(m_q,m_q)$, $\hat F_{K,K^*}=\hat F_{PS,V}(m_q,m_s)$). 
The physical values of the decay constants are then obtained by using  
\be f_D = \hat R_H(m_{charm},m_q) \times f_\pi^{(exp.)}\, , \quad 
f_{D_s} = \hat R_{H_s}(m_{charm},m_s) \times f_K^{(exp.)}\, ,\ee
and similarly for the vector mesons
\be f_{D^*} = \hat R_{H^*}(m_{charm},m_q) \times f_\rho^{(exp.)}\, , \quad 
f_{D^*_s} = \hat R_{H^*_s}(m_{charm},m_s) \times f_{K^*}^{(exp.)}\, .\ee
The experimental values of the decay constants that we use are the following ones~\cite{experiment}: $f_\pi^{(exp.)}= 131\,\mev$, $f_K^{(exp.)}=160\,\mev$, $f_\rho^{(exp.)}=208(2)\,\mev$, $f_{K^*}^{(exp.)}=214(7)\,\mev$.

The results are given in Tab.~\ref{tab:fd}. We also give the decay constants obtained by including the
KLM factor which we discuss in Sec.~\ref{sec:bmesons}. The differences can be used for an estimate of the residual ${\cal O}(a^2)$, discretization
errors in the determination of the matrix elements. In Tab.~\ref{tab:fds}, we list the corresponding results for vector mesons.
\begin{table}[h] 
\centering
\begin{tabular}{|c|cc|cc|} \hline 
{\phantom{\huge{l}}}\raisebox{-.2cm}{\phantom{\huge{j}}}
{ }&  \multicolumn{2}{c|}{$f_{D_q}$ [MeV]}  &   
\multicolumn{2}{c|}{$f_{D_s}$ [MeV]} \\ \cline{2-5}
{\phantom{\huge{l}}}\raisebox{-.2cm}{\phantom{\huge{j}}}
{ No KLM factor}   & {\rm linear in $1/M_H$} & {\rm quad. in $1/M_H$} &  {\rm linear in $1/M_{H_s}$} & {\rm quad. in $1/M_{H_s}$}  \\ \hline 
{\phantom{\huge{l}}}\raisebox{-.2cm}{\phantom{\huge{j}}}
{\phantom{\huge{l}}}\raisebox{-.2cm}{\phantom{\huge{j}}}
{\hspace*{-.7cm} $f_{D_\ell}$ using  $\hat{F}_{\pi}$} & 201(22) & 200(21) &  -- & -- \\ 
{\phantom{\huge{l}}}\raisebox{-.2cm}{\phantom{\huge{j}}}
{ $f_{D_\ell}$ using $\hat{F}_{K}$} & -- & -- &  239(18) & 238(16)\\ 
{\phantom{\huge{l}}}\raisebox{-.2cm}{\phantom{\huge{j}}}
{ $f_{D_\ell}$ using $a^{-1}(m_{K^*})$ }& {\sf 213(14)} &{\sf 212(15)} &  {\sf 233(11)} &{\sf 232(12)}
\\ \hline  \hline 
{\phantom{\huge{l}}}\raisebox{-.2cm}{\phantom{\huge{j}}}
{ } &  \multicolumn{2}{c|}{$f_{D_q}$ [MeV]}  &   
\multicolumn{2}{c|}{$f_{D_s}$ [MeV]} \\ \cline{2-5}
{\phantom{\huge{l}}}\raisebox{-.2cm}{\phantom{\huge{j}}}
{ With KLM factor}   & {\rm linear in $1/M_H$} & {\rm quad. in $1/M_H$} &  {\rm linear in $1/M_{H_s}$} & {\rm quad. in $1/M_{H_s}$}  \\ \hline 
{\phantom{\huge{l}}}\raisebox{-.2cm}{\phantom{\huge{j}}}
{\phantom{\huge{l}}}\raisebox{-.2cm}{\phantom{\huge{j}}}
{\hspace*{-.7cm} $f_{D_\ell}$ using  $\hat{F}_{\pi}$} & 199(22) & 198(21) &  -- & -- \\ 
{\phantom{\huge{l}}}\raisebox{-.2cm}{\phantom{\huge{j}}}
{ $f_{D_\ell}$ using $\hat{F}_{K}$} & -- & -- &  237(17) & 236(16)\\ {\phantom{\huge{l}}}\raisebox{-.2cm}{\phantom{\huge{j}}}
{ $f_{D_\ell}$ using $a^{-1}(m_{K^*}) $ }& {\sf 211(14)} &{\sf 210(15)} &  {\sf 231(12)} &{\sf 230(13)}
\\ \hline 
\end{tabular}
\caption{{\it Pseudoscalar decay constants for $D$-mesons using the scaling law in Eq.~{\rm (\protect\ref{eq:scaling})}. Results
including the KLM factor discussed in the text, are given in the lower part of the table.}}
\label{tab:fd}
\end{table}
\begin{table}[h] 
\centering
\begin{tabular}{|c|cc|cc|}  \hline
{\phantom{\huge{l}}}\raisebox{-.2cm}{\phantom{\huge{j}}}
{ } &  \multicolumn{2}{c|}{$f_{D^*_q}$ [MeV]}  &   
\multicolumn{2}{c|}{$f_{D^*_s}$ [MeV]} \\ \cline{2-5}
{\phantom{\huge{l}}}\raisebox{-.2cm}{\phantom{\huge{j}}}
{ No KLM factor}   & {\rm linear in $1/M_H$} & {\rm quad. in $1/M_H$} &  {\rm linear in $1/M_{H_s}$} & {\rm quad. in $1/M_{H_s}$}  \\ \hline 
{\phantom{\huge{l}}}\raisebox{-.2cm}{\phantom{\huge{j}}}
{\phantom{\huge{l}}}\raisebox{-.2cm}{\phantom{\huge{j}}}
{\hspace*{-.7cm} $f_{D^*_\ell}$ using  $\hat{F}_{\rho}$} & 246(30) & 244(32) &  -- & -- \\ 
{\phantom{\huge{l}}}\raisebox{-.2cm}{\phantom{\huge{j}}}
{ $f_{D^*_\ell}$ using $\hat{F}_{K^*}$} & -- & -- &  255(17) & 253(18)\\ 
{\phantom{\huge{l}}}\raisebox{-.2cm}{\phantom{\huge{j}}}
{ $f_{D^*_\ell}$ using $a^{-1}(m_{K^*})$ }& {\sf 248(19)} &{\sf 246(21)} &  {\sf 275(15)} &{\sf 273(16)}
\\ \hline \hline
{\phantom{\huge{l}}}\raisebox{-.2cm}{\phantom{\huge{j}}}
{ } &  \multicolumn{2}{c|}{$f_{D^*_q}$ [MeV]}  &   
\multicolumn{2}{c|}{$f_{D^*_s}$ [MeV]} \\ \cline{2-5}
{\phantom{\huge{l}}}\raisebox{-.2cm}{\phantom{\huge{j}}}
{ With KLM factor}   & {\rm linear in $1/M_H$} & {\rm quad. in $1/M_H$} &  {\rm linear in $1/M_{H_s}$} & {\rm quad. in $1/M_{H_s}$}  \\ \hline 
{\phantom{\huge{l}}}\raisebox{-.2cm}{\phantom{\huge{j}}}
{\phantom{\huge{l}}}\raisebox{-.2cm}{\phantom{\huge{j}}}
{\hspace*{-.7cm} $f_{D^*_\ell}$ using  $\hat{F}_{\rho}$} & 243(31) & 241(33) &  -- & -- \\ 
{\phantom{\huge{l}}}\raisebox{-.2cm}{\phantom{\huge{j}}}
{ $f_{D^*_\ell}$ using $\hat{F}_{K^*}$} & -- & -- &  252(18) & 251(19)\\ 
{\phantom{\huge{l}}}\raisebox{-.2cm}{\phantom{\huge{j}}}
{ $f_{D^*_\ell}$ using $a^{-1}(m_{K^*})$ }& {\sf 245(20)} &{\sf 243(21)} &  {\sf 272(16)} &{\sf 270(16)}
\\ \hline 
\end{tabular}
\caption{{\it Vector decay constants for charmed-mesons obtained by using the scaling law in Eq.~{\rm (\protect\ref{eq:scaling})}. Results,
including the KLM factor discussed in the text, are also given.}}
\label{tab:fds}
\end{table}
\par
\begin{figure}
\vspace{-.5cm}
\centering
\hspace*{.3cm} \epsfbox{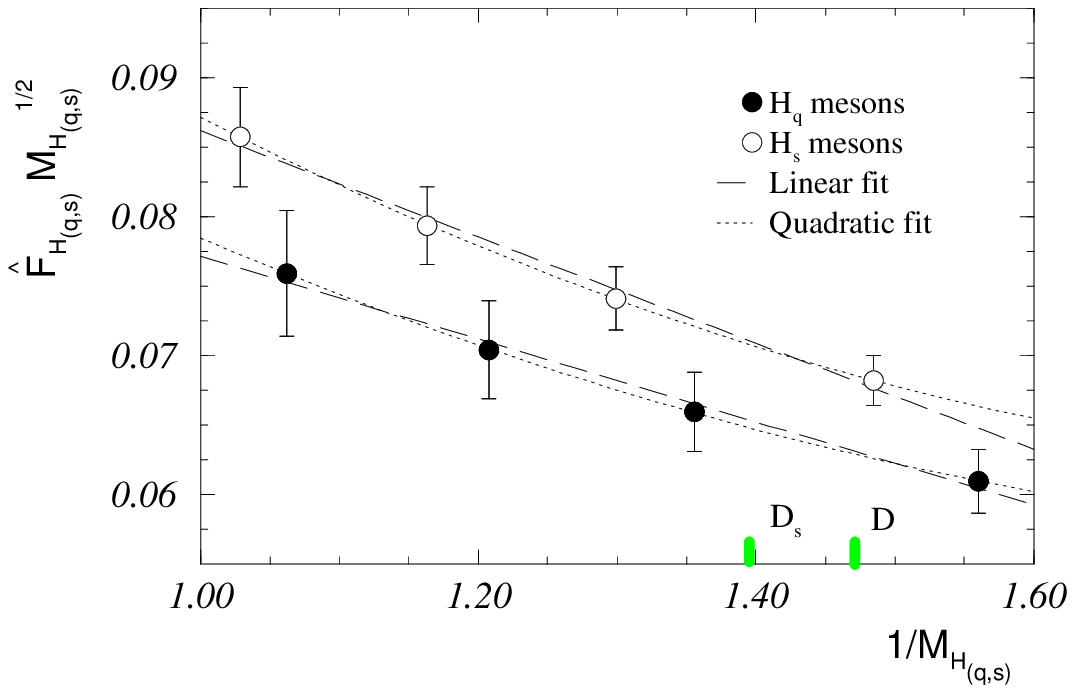}
\vspace*{0.5cm} \hspace*{0.3cm}\epsfbox{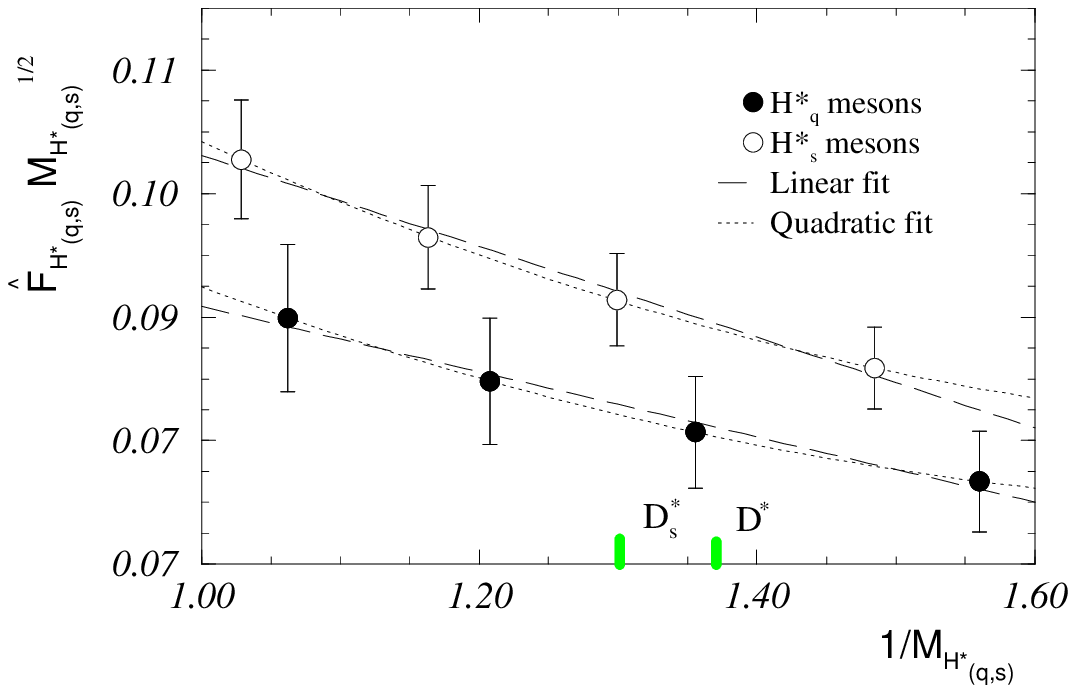}
\caption{\it Results of the linear and quadratic fits for pseudoscalar and vector  mesons, in lattice units. The gray dashed lines correspond to $f_D$ ($f_{D_s}$), and $f_{D^*}$ ($f_{D_s^*}$) respectively.}
\label{fig:lqfit}
\end{figure}

 Whether we use a linear or a quadratic fit to interpolate to $\kappa_{charm}$, our  results in the $D$-sector remain practically unchanged. In order to illustrate the stability of the results for $D$ mesons, we also show in Fig.~\ref{fig:lqfit}, the results of the linear and quadratic fits in $1/M_H$.  

\par The observed stability makes these results quite remarkable: we use the non-perturbatively improved action; the operators and the renormalization constants are also improved; the results obtained by using the heavy quark scaling laws are unchanged, regardless of whether we take quadratic ($1/M_H^2$) corrections into account or not; the results are practically insensitive to the presence of KLM factors; there is no important dependence on the quantity chosen to fix the physical normalization. The errors that we quoted in (\ref{eq:resultd}), are obtained in the following way: 
a central value is fixed by the result obtained from the linear fit in $1/M_H$,  with the scale fixed by $m_{K^*}$, and the KLM factor included; we quote the statistical error as estimated using the jackknife procedure; all the residual differences are lumped into the systematic uncertainty (the difference between the central values of the results obtained by using different quantities for the scale fixing, and the difference with the central value of the result obtained from the quadratic fit in $1/M_H$). It is also worth  noticing the remarkable stability of the ratio ${f_{D_s}/f_{D}}$~(see Eq.~(\ref{eq:resultd})), although it may be questioned whether we are really able to predict the SU(3) breaking properly in the quenched approximation. More discussion on this point will be given in Sec.~\ref{sec:scaling}.
\section{$B$ mesons}
\label{sec:bmesons}
In this section, we present the results of the {\sl extrapolation} of the decay constants to the $B$ mesons, and discuss the discretization errors in the
extrapolation.
\par  When extrapolating the raw data obtained for $m_h \sim m_{charm}$
to the $B$-sector, two important effects may arise. On the one hand, the inclusion of the quadratic term  $\Phi_2/M^2_H$ in Eq.~(\ref{eq:scaling})
may change appreciably the results of the linear fit, on the other the ${\cal O}(a)$ corrections (${\cal{O}}(a)$ terms proportional to $c_{V,A}$ and $b_{V,A}$ in (\ref{eq:rco})) become much larger. This is to be contrasted to the case of $D$-mesons, where the inclusion of the quadratic corrections leaves the results essentially unchanged, cf. Tabs.~\ref{tab:fd} and \ref{tab:fds}. 
\par  The effect of  $c_{A}$, $c_{V}$, $b_{V}$ and $b_{A}$ is sizable for the scaling behaviour of $f_{B,B^*}$. Note also that if we used $c_{V}^{\scr {NP}}$, this effect would be huge for the vector decay constant. 
For instance, in the range of quark masses considered in our simulation, the renormalization constants $Z_{V,A}(m)$, defined in Eq.~(\ref{eq:rco}), increase by $20\div 50\,\%$, relatively to their values in the chiral limit.
Since $Z_{V}(m)$ and $Z_{A}(m)$ are multiplicative factors, their effect 
is very important for the extrapolation to the $B$-sector. This is illustrated in Fig.~\ref{fig:fig011}: 
\begin{figure}
\centering
\hspace*{0.3cm}\epsfbox{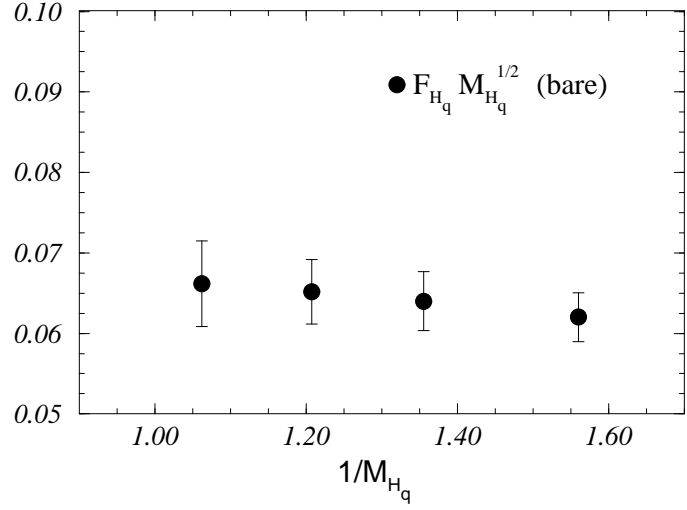}
\vspace*{0.5 cm} \hspace*{0.3cm}\epsfbox{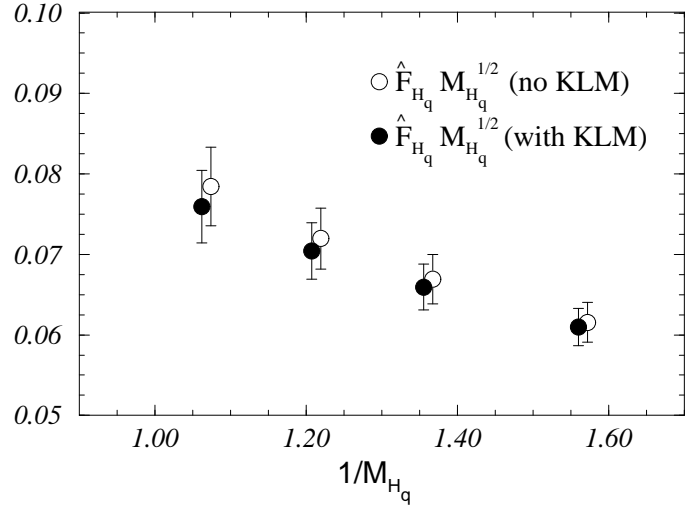}
\caption{\it Heavy-light pseudoscalar decay constant as a function of $1/M_H$. 
The two figures illustrate the influence of the renormalization constant $Z_A(m)$ 
on the $1/M_H$ dependence of the decay constant.}
\label{fig:fig011}
\end{figure} 
when $Z_A(m)$ is included, we note that the quadratic fit is more desired, although  the linear one is compatible with the data. 
The embarrassing point is that the values of $f_B$ and $f_{B^*}$, as obtained from the linear and quadratic fit, are hardly compatible, see Tabs.~\ref{tab:fb} and \ref{tab:fbs}. This is particularly pronounced for $B_s^{(*)}$ mesons.
\begin{table}[h] 
\centering
\begin{tabular}{|c|cc|cc|} \hline 
{\phantom{\huge{l}}}\raisebox{-.2cm}{\phantom{\huge{j}}}
{ } &  \multicolumn{2}{c|}{$f_{B_q}$ [MeV]}  &   
\multicolumn{2}{c|}{$f_{B_s}$ [MeV]} \\ \cline{2-5}
{\phantom{\huge{l}}}\raisebox{-.2cm}{\phantom{\huge{j}}}
{ No KLM factor}   & {\rm linear in $1/M_H$} & {\rm quad. in $1/M_H$} &
  {\rm linear in $1/M_{H_s}$} & {\rm quad. in $1/M_{H_s}$}  \\ \hline 
{\phantom{\huge{l}}}\raisebox{-.2cm}{\phantom{\huge{j}}}
{\phantom{\huge{l}}}\raisebox{-.2cm}{\phantom{\huge{j}}}
{\hspace*{-.7cm} $f_{B_\ell}$ using  $\hat{F}_{\pi}$} & 176(19) & 208(27) &  -- & -- \\ 
{\phantom{\huge{l}}}\raisebox{-.2cm}{\phantom{\huge{j}}}
{ $f_{B_\ell}$ using $\hat{F}_{K}$} & -- & -- &  217(14) & 255(20)\\ 
{\phantom{\huge{l}}}\raisebox{-.2cm}{\phantom{\huge{j}}}
{ $f_{B_\ell}$ using $a^{-1}(m_{K^*})$ }& {\sf 187(19)} &{\sf 220(25)} &  {\sf 212(16)} &{\sf 249(20)}
\\ \hline \hline
{\phantom{\huge{l}}}\raisebox{-.2cm}{\phantom{\huge{j}}}
{ } &  \multicolumn{2}{c|}{$f_{B_q}$ [MeV]}  &   
\multicolumn{2}{c|}{$f_{B_s}$ [MeV]} \\ \cline{2-5}
{\phantom{\huge{l}}}\raisebox{-.2cm}{\phantom{\huge{j}}}
{ With KLM factor}   & {\rm linear in $1/M_H$} & {\rm quad. in $1/M_H$} &  {\rm linear in $1/M_{H_s}$} & {\rm quad. in $1/M_{H_s}$}  \\ \hline 
{\phantom{\huge{l}}}\raisebox{-.2cm}{\phantom{\huge{j}}}
{\phantom{\huge{l}}}\raisebox{-.2cm}{\phantom{\huge{j}}}
{\hspace*{-.7cm} $f_{B_\ell}$ using  $\hat{F}_{\pi}$} & 170(18) & 193(25) &  -- & -- \\ 
{\phantom{\huge{l}}}\raisebox{-.2cm}{\phantom{\huge{j}}}
{ $f_{B_\ell}$ using $\hat{F}_{K}$} & -- & -- &  209(13) & 238(19)\\ {\phantom{\huge{l}}}\raisebox{-.2cm}{\phantom{\huge{j}}}
{ $f_{B_\ell}$ using $a^{-1}(m_{K^*}) $ }& {\sf 179(18)} &{\sf 205(24)} &  {\sf 204(16)} &{\sf 232(19)}
\\ \hline 
\end{tabular}
\caption{{\it Pseudoscalar decay constants for $B$-mesons using the scaling law in Eq.~{\rm (\protect\ref{eq:scaling})}.
 Results with the KLM factor included are listed in lower part of the table.}}
\label{tab:fb}
\end{table}
\begin{table}[h]
\centering
\begin{tabular}{|c|cc|cc|} \hline 
{\phantom{\huge{l}}}\raisebox{-.2cm}{\phantom{\huge{j}}}
{ } &  \multicolumn{2}{c|}{$f_{B^*_q}$ [MeV]}  &   
\multicolumn{2}{c|}{$f_{B^*_s}$ [MeV]} \\ \cline{2-5}
{\phantom{\huge{l}}}\raisebox{-.2cm}{\phantom{\huge{j}}}
{ No KLM factor}   & {\rm linear in $1/M_H$} & {\rm quad. in $1/M_H$} &  {\rm linear in $1/M_{H_s}$} & {\rm quad. in $1/M_{H_s}$}  \\ \hline 
{\phantom{\huge{l}}}\raisebox{-.2cm}{\phantom{\huge{j}}}
{\phantom{\huge{l}}}\raisebox{-.2cm}{\phantom{\huge{j}}}
{\hspace*{-.7cm} $f_{B^*_\ell}$ using  $\hat{F}_{\rho}$} & 204(34) & 239(39) &  -- & -- \\ 
{\phantom{\huge{l}}}\raisebox{-.2cm}{\phantom{\huge{j}}}
{ $f_{B^*_\ell}$ using $\hat{F}_{K^*}$} & -- & -- &  222(22) & 260(25)\\ 
{\phantom{\huge{l}}}\raisebox{-.2cm}{\phantom{\huge{j}}}
{ $f_{B^*_\ell}$ using $a^{-1}(m_{K^*})$ }& {\sf 205(25)} &{\sf 241(32)} &  {\sf 239(21)} &{\sf 280(24)}
\\ \hline \hline
{\phantom{\huge{l}}}\raisebox{-.2cm}{\phantom{\huge{j}}}
{ } &  \multicolumn{2}{c|}{$f_{B^*_q}$ [MeV]}  &   
\multicolumn{2}{c|}{$f_{B^*_s}$ [MeV]} \\ \cline{2-5}
{\phantom{\huge{l}}}\raisebox{-.2cm}{\phantom{\huge{j}}}
{ With KLM factor}   & {\rm linear in $1/M_H$} & {\rm quad. in $1/M_H$} &  {\rm linear in $1/M_{H_s}$} & {\rm quad. in $1/M_{H_s}$}  \\ \hline 
{\phantom{\huge{l}}}\raisebox{-.2cm}{\phantom{\huge{j}}}
{\phantom{\huge{l}}}\raisebox{-.2cm}{\phantom{\huge{j}}}
{\hspace*{-.7cm} $f_{B^*_\ell}$ using  $\hat{F}_{\rho}$} & 194(32) & 225(37) &  -- & -- \\ 
{\phantom{\huge{l}}}\raisebox{-.2cm}{\phantom{\huge{j}}}
{ $f_{B^*_\ell}$ using $\hat{F}_{K^*}$} & -- & -- &  213(22) & 241(24)\\ 
{\phantom{\huge{l}}}\raisebox{-.2cm}{\phantom{\huge{j}}}
{ $f_{B^*_\ell}$ using $a^{-1}(m_{K^*})$ }& {\sf 196(24)} &{\sf 227(30)} &  {\sf 229(20)} &{\sf 260(23)}
\\ \hline 
\end{tabular}
\caption{{\it Vector decay constants for $B$-mesons using the scaling law in Eq.~{\rm(\protect\ref{eq:scaling})}.
Results including the KLM factor discussed in the text, are also given.}}
\label{tab:fbs}
\end{table}

\par The curvature in the fit to $f_B$ could partially be induced by ${\cal O}(a^2)$ terms, still present
in the calculation of the matrix elements.  
A possible way to account for some of these effects is through the so-called  KLM factor~\cite{klm}. 
In our case, this means that, besides the factor $(1 \,+ \,b_J\, ma)$ already included in the definition of the
renormalized currents~(\ref{eq:rco}), we may try to include the effects of  higher order terms in $ma$, 
by using the following relation
\bea
Z_J (m_h, m_\ell) &=& Z_J(0) \, \left[ {\sqrt{1 + am_h}\,\sqrt{1 + am_\ell}\over 1 \,+\, am}\right]
\left( 1 + \  b_J\, am\right) \cr
{\phantom{\Huge{l}}}\raisebox{+.1cm}{\phantom{\Huge{j}}}
   &\simeq& Z_J(0)\, \left( 1 + \  b_J\, am \right) + {\cal O}(a^2), 
\label{eq:33a}
\eea
where $am=(am_h + am_\ell)/2$, and $am_i$ is the usual expression for the bare quark mass (\ref{eq:qmasses}).
Equation~(\ref{eq:33a}) is a consequence of the redefinition of a quark field, $q\to \sqrt{1+am}\; q$ 
(in the KLM way), which comes from 
the comparison of the {\em free} lattice quark propagator to its continuum counterpart. The results which include the
KLM correction are given in the lower part of Tabs.~\ref{tab:fd}, \ref{tab:fds}, \ref{tab:fb} and {\ref{tab:fbs}. 
In the case of $D$-mesons, the effect of KLM is indeed negligible. In the case of $B$-mesons, we 
observe a slight change in the central values, {\it e.g.} $f_B= 187 $ MeV $ \to 179 $ MeV. However, the distance between
the  values obtained with linear and with quadratic fits remains essentially unchanged.
In the absence of a larger  range of masses, we are unable to reduce the difference between results obtained
with the linear and quadratic
fits. As it has been done for $D$-mesons, we quote the results of the linear fits as our central values, and include in the systematic error the
differences between our central values and~$i$)~the results from the quadratic fit; $ii$) the results without the KLM factor incorporated; $iii$) the results obtained by using other quantities ($f_K,f_\pi$) to extract the physical values. Our final results are those given in Eq.~(\ref{eq:fbl}).
\section{Scaling laws and related issues}
\label{sec:scaling}
In this section, we discuss several interesting quantities for the study of 
the scaling laws predicted by the HQET, and their validity in the range of quark masses between the charmed and the bottom one.
We introduce several ratios of decay constants which are useful to get some physical information.
\par We first consider the scaling law  for the decay constants.
The results for the coefficients in Eq.~(\ref{eq:scaling}), as obtained from our fits, are given in {Tab.~10.} To translate these coefficients into physical units, we have used  $a^{-1}(m_{K^*})$.
\begin{table}[h!!]
\begin{center}
\begin{tabular}{|c||cc|cc|}  \hline 
{\rm fit}  & \multicolumn{2}{c|}{$m_\ell=m_{q}$ }  &   \multicolumn{2}{c|}{$m_\ell=m_{s}$ } \\
{\rm parameters}& {\rm linear} & {\rm quadratic} &  {\rm linear} & {\rm quadratic}  \\ \hline  \hline {\phantom{\Huge{l}}}\raisebox{-.2cm}{\phantom{\Huge{j}}}
$\Phi_0^{PS}\,{\rm [GeV^{3/2}]}$ & 0.48(5) & 0.66(11)  &0.56(5) & 0.74(8) \\
{\phantom{\Huge{l}}}\raisebox{-.2cm}{\phantom{\Huge{j}}}
$\Phi_1^{PS}/\Phi_0^{PS}\,{\rm [GeV]}$ &-0.75(6) &-1.60(22)  &-0.83(5) &-1.70(16)\\
{\phantom{\Huge{l}}}\raisebox{-.2cm}{\phantom{\Huge{j}}}
$\sqrt{\Phi_2^{PS}/\Phi_0^{PS}}\,{\rm [GeV]}$ & -- &1.03(8) &-- &1.08(6)\\ \hline 
{\phantom{\Huge{l}}}\raisebox{-.2cm}{\phantom{\Huge{j}}}
$\Phi_0^{V}\,{\rm [GeV^{3/2}]}$ & 0.51(7) & 0.70(12)  &0.61(6) & 0.81(10) \\
{\phantom{\Huge{l}}}\raisebox{-.2cm}{\phantom{\Huge{j}}}
$\Phi_1^{V}/\Phi_0^{V}\,{\rm [GeV]}$ &-0.63(9) &-1.62(24)  &-0.74(6) &-1.65(17)\\
{\phantom{\Huge{l}}}\raisebox{-.2cm}{\phantom{\Huge{j}}}
$\sqrt{\Phi_2^{V}/\Phi_0^{V}}\,{\rm [GeV]}$ & -- &1.09(9) &-- &1.08(6)\\ \hline
\end{tabular}
{\caption{\it Fit parameters in physical units for pseudoscalar {\rm($PS$)} and vector {\rm($V$)} heavy-light mesons.}}
\end{center}
\label{coeffs1}
\end{table}
\noindent
The leading term from the linear fit, $\Phi_0= 0.48(5)\,GeV^{3/2}$, is in  good  agreement with the findings of previous studies~\cite{bernard,ape,jlqcd,ukqcd}. We also note that this value is compatible with  the results of QCD sum rules~\cite{heavyqcdsr}, $\Phi_0=(0.4\div 0.6)\,\gev^{3/2}$, when the large perturbative QCD  corrections are included~\footnote{ Without these corrections, the result would be $\Phi_0=0.30(5)$~\cite{heavyqcdsr}.}.

\par Following the Refs.~\cite{bmasc,ukqcd}, we now consider the spin scaling relation on the lattice:
\bea
U({\overline{M}_H})={f_H\over f_{H^*}} = \xi_0 + {\xi_1\over {\overline{M}_H}}+ {\xi_2\over {\overline{M}_H^2}} \ , 
\eea
where $\overline{M}_H = (3 M_{H^*} + M_H)/4$  is the spin averaged mass (which we already used in Sec.~\ref{sec:dmesons}), and $\xi_{0,1,2}$ are parameters which we obtain  by fitting our data. From heavy quark symmetry, one expects that $\xi_0 = 1$ (up to logarithmic corrections). 
For completeness, we tabulated $\overline{M}_H$ and $f_H/f_{H^*}$ in Tab.~\ref{tab:UUU}.
\begin{table}[h]
\vspace*{5mm}
\centering
\begin{tabular}{|c|c c c c|} \hline 
{\phantom{\Huge{l}}}\raisebox{-.2cm}{\phantom{\Huge{j}}}
{ \rm `heavy flavor'} & { $\kappa_{h1}$}& { $\kappa_{h2}$} & {  $\kappa_{h3}$}& { $\kappa_{h4}$}\\ \hline \hline
{\phantom{\Huge{l}}}\raisebox{-.2cm}{\phantom{\Huge{j}}}
{$\overline{M}_H = (3 M_{H^*}+M_{H})/4$}& 0.6705(48) & 0.7620(53)& 0.8486(56)& 0.9579(61) \\ 
{\phantom{\Huge{l}}}\raisebox{-.2cm}{\phantom{\Huge{j}}}
{$\overline{M}_{H_s} = (3 M_{H^*_s}+M_{H_s})/4$}& 0.7028(58) & 0.7940(58)& 0.8802(58)& 0.9890(60) \\ \hline
{\phantom{\Huge{l}}}\raisebox{-.2cm}{\phantom{\Huge{j}}}
{$U(\overline{M}_H) = f_{H}/f_{H^*}$}   &  0.851(30) & 0.869(29)& 0.879(29)& 0.888(31) \\ {\phantom{\Huge{l}}}\raisebox{-.2cm}{\phantom{\Huge{j}}}
{$U(\overline{M}_{H_s}) = f_{H_s}/f_{H^*_s}$}   &  0.844(22) & 0.858(21)& 0.867(21)& 0.876(22) \\ \hline 
\end{tabular}
\caption{\it Spin averaged masses and ratios of pseudoscalar and vector decay constants. For $\overline{M}_H$, the light quark mass $q=u,d$, is understood.}
\label{tab:UUU}
\end{table}
The results of our fits in physical units, are
\bea
&(lin.)& \xi_0 =0.997(68)\, ,  \quad \xi_1/\xi_0 = -0.23(11)\, GeV\, ,\nonumber \\
&(quad.)& \xi_0 =0.89(12)\, , \quad \xi_1/\xi_0 = 0.17(49)\, GeV\, , \quad \sqrt{\xi_2/\xi_0}=-0.67(18)\, GeV \, ,
\label{eq:csiU} \eea
where the physical values were obtained by using $a^{-1}(m_{K^*})$.
Data points, and extrapolated values, are   displayed in Fig.~\ref{fig:fig14}.
\begin{figure}\centering
\hspace*{0.3 cm} \epsfbox{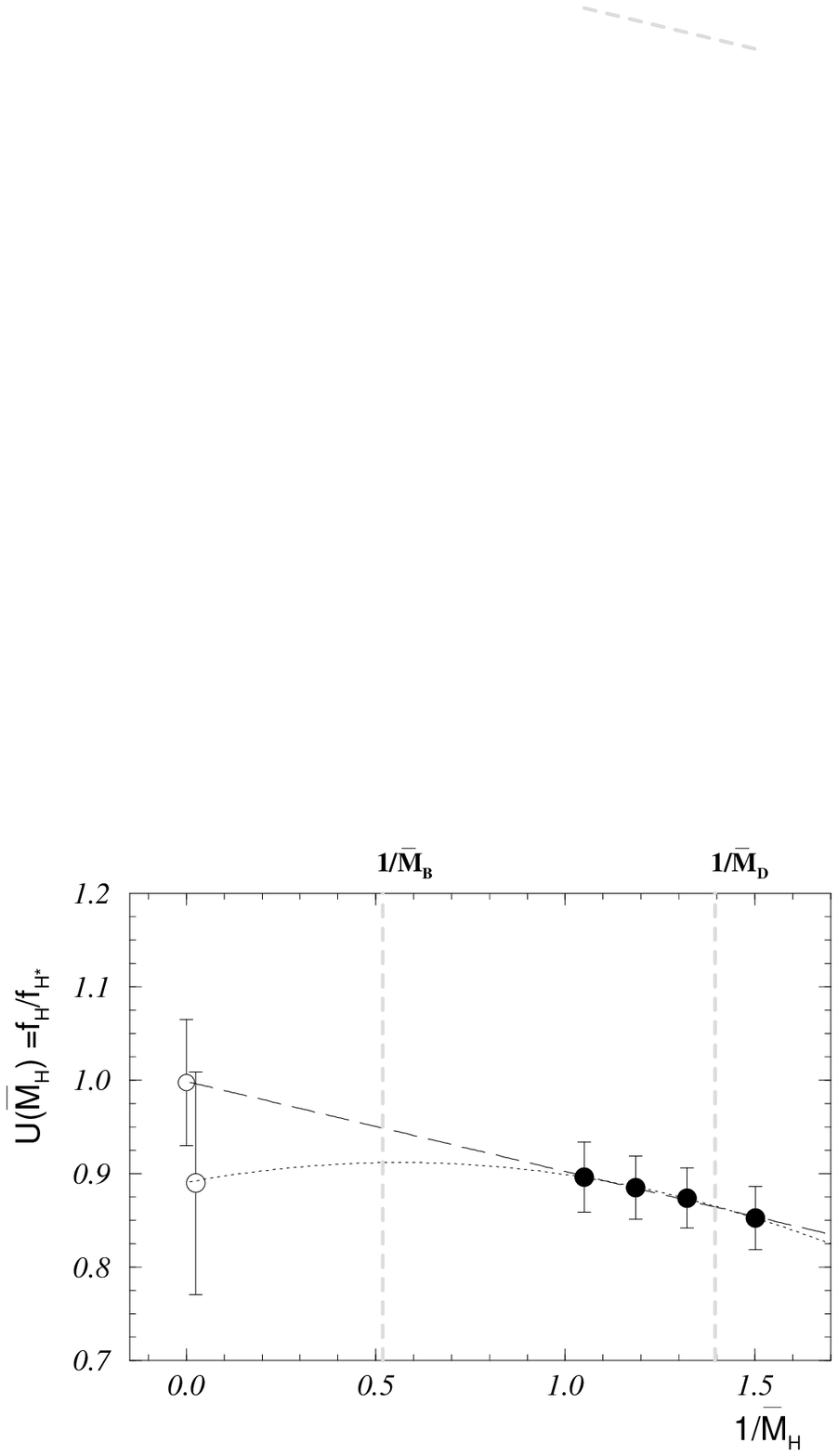}
\caption{\it Ratio of the heavy-light decay constants. 
The linear fit approaches very well the expected asymptotic value $U(\overline{M}_H \to \infty)= \xi_0= 1$. The results refer to mesons with the light quark extrapolated to $q=u,d$.}
\label{fig:fig14}
\end{figure}
We see that the  scaling law is very well satisfied by using the linear fit. The inclusion of the quadratic term, although irrelevant in the directly accessible region of the meson masses, produces large deviation
from the expected extrapolated value  $\xi_0=1$, as $\overline{M}_H \to \infty$. Thus, by using the linear fit, we arrive at 
\bea
U(\overline{M}_D) = 0.860(28) , \quad \;
U(\overline{M}_B) = 0.933(47) , 
\eea
and
\bea
U(\overline{M}_{D_s}) = 0.868(21) ,  \quad \; 
U(\overline{M}_{B_s}) = 0.915(33).
\eea

\par
We end this section by presenting a set of ratios which may be explored in order to extract the physical decay constants by using some measured quantities.

\begin{itemize}
\item 
As it was suggested in Ref.~\cite{bernard},
the decay constants can be conveniently presented in terms of $f_{D_s}$, which is already 
measured~\footnote{$f_{D_s}^{(exp.)}=254\pm 31\, \mev$~\protect\cite{rich}.}:
\bea  
{ f_{B}\over f_{D_s}} = 0.78\pm 0.04\,^{+11}_{-0}; &\quad& 
{ f_{B_s}\over f_{D_s}}  = 0.88\pm 0.03\,^{+12}_{-0}; \nonumber  \\
{ f_{D^*}\over f_{D_s}} =  1.06\pm 0.05; &\quad&
{ f_{D^*_s}\over f_{D_s}} =  1.17\pm 0.03; \nonumber  \\
{ f_{B^*}\over f_{D_s}} = 0.85\pm 0.07\,^{+13}_{-0}&\quad&{ f_{B^*_s}\over f_{D_s}} = 0.99\pm 0.05\,^{+14}_{-0}.
\eea
The error estimates are obtained in the same way as in Sec.~\ref{sec:bmesons}. $f_B/f_{D_s}$ is the value which has been used in Eq.~(\ref{aka}), as an alternative way to extract the value for $f_B$.
\item 
Other phenomenologically interesting ratios for testing the factorization hypothesis in non-leptonic modes, are (see~\cite{NeubertQCD97}):
\bea
{ f_{D^*_s}\over f_{\rho}} = 1.29(14);\,\quad \,
{ f_{D_s}\over f_{\pi}} = 1.66(19). 
\eea
\item In the quenched approximation, the $SU(3)$-breaking parameter $r_K-1 \equiv f_K/f_\pi - 1$, is expected to be smaller
than its experimental value. A smaller value of $r_K-1$ is predicted by one-loop quenched chiral perturbation theory~\cite{golterman}, and is verified in numerous simulations (with either unimproved or improved actions and operators~\cite{spectrum}). A similar effect is also expected for $r_H=f_{H_s}/f_{H}$ ($r_D=f_{D_s}/f_D$, $r_B=f_{B_s}/f_{B}$)~\cite{zhang}.
In this respect (in the hope of reducing the quenching errors), it may be interesting to examine the Grinstein-type double ratio ${\frak{R}}_H=r_H/r_D$~\cite{gr}. From our data, we have
\be
{\frak{R}}_{H_{h_1}}= 0.995(3)\, ,  \quad  {\frak{R}}_{H_{h_2}}= 1.003(4)\, , \quad  {\frak{R}}_{H_{h_3}}= 1.009(6)\,
 , \quad {\frak{R}}_{H_{h_4}}= 1.014(9)\ , \nn
\ee
which upon an extrapolation to  the B-meson mass, amounts to 
\be 
{\frak{R}}_{B}^{(lin)}= 1.035(17) \ , \quad {\frak{R}}_{B}^{(quad)}= 1.028(33)  \ .
\ee
Using ${\frak{R}}_{B}^{(lin)}$ and $r_D=1.10(2)$ from (\ref{eq:resultd}), we have  $r_B= 1.134(34)$, in perfect agreement with the direct determination, given in (\ref{eq:fbl}).
\item  The double ratio can be used to estimate the quenching 
    errors in the predicted values of $r_D$ and $r_B$. To this purpose, we define
 \be \bar r_H = \frac{r_H}{r_K} \left(\frac{f_K}{f_\pi}\right)^{(exp.)} \ ,\ee
where $r_H$ and $r_K$, are obtained in the quenched lattice calculation. Using our data ($r_K=1.12(5)$~\cite{spectrum}) and
$(f_K/f_\pi)^{(exp.)}=1.22$~\cite{experiment}, we end up with
\bea
\bar r_{D}^{(lin.\& quad.)}&=& 1.19(5);\\
\bar r_{B}^{(lin. \& quad.)}&=& 1.23(6),
\label{su3}
\eea
which are $\sim 9\%$ larger than the results obtained directly and quoted in Eqs.~{(\ref{eq:resultd}) and (\ref{eq:fbl})}. If this difference of $9\%$ is the realistic estimate of the quenching errors, they are much smaller than the pessimistic estimate of Ref.~\cite{zhang}, where $\sim 20\%$ of (quenching) error was predicted. Note that the ratios $\bar r_{D}$ and $\bar r_{B}$, do not depend on the fit we use~(linear or quadratic). A similar game with $f_B/f_{D_s}$ results in 
\bea
{f_B\over f_{D_s}} = 0.71\pm 0.04\,{}^{+10}_{-0}
\eea
which gives $f_B=\: (180\pm 26({\rm exp.}) \,^{+29}_{-10}({\rm theo.}) )\; {\rm MeV}$, where we accounted for the experimental value for $f_{D_s}^{(exp.)}$. This result agrees with the value we reported in~(\ref{eq:fbl}). 

\end{itemize}
\newpage
\section{Conclusion}
We have analyzed masses and decay constants of heavy-light pseudoscalar and vector mesons, using the non-perturbatively improved action and currents. 
Particular attention has been paid to the errors coming from the extrapolation in the light and heavy quark masses.

 We find that the  hyperfine splitting is definitely below
the experimental value, in spite of the improved action.

The values predicted for the decay constants of $D$ mesons are extremely stable against variations of the fitting procedure, inclusion of KLM factors etc. Thus we believe that the main error on these quantities is the quenching error.

On the contrary, we find larger uncertainties for the $B$-meson decay constants, mainly due to the amplification of discretization effects when extrapolating
to the $b$-quark mass, and to the uncertainty in the extrapolation procedure.
In spite of these uncertainties, and of the fact that our results are obtained at a single value of the lattice spacing, we believe a value of $f_B$ much lower than {$170\, \mev$} rather unlikely. Indeed, for
$\beta \ge 6.0$, with Wilson-like fermions at fixed lattice spacing, almost all lattice calculations give $f_B$ larger than $160$ MeV. This value has been quoted as the ``world average" obtained in Ref.~\cite{draper98}, after combining results obtained with propagating quarks, with those obtained using some effective theory, as NRQCD~\cite{ali-khan}, or the FNAL action~\cite{fnal}.
Low values of $f_B$ with propagating quarks are obtained only after extrapolating in $a$ to the continuum
limit~\cite{bernard,draper98}, with procedures which we believe are questionable (for example by including data at low values of $\beta$, {\it i.e.}
too close to the strong coupling regime). Our results, which should have smaller discretization
errors than other calculations at fixed lattice spacing, confirm a value of $f_B$ (in the quenched approximation) larger than
 $170\,\mev$.
A (rather) indirect evidence that a larger value of $f_B$ is preferred can be obtained by combining
$f_B/f_{D_s}$ from the lattice with the experimental value of $f_{D_s}$. This gives  $f_B\simeq 180\div 190$ MeV,
with an error of about $40$ MeV. 
Finally, we used the Grinstein double-ratio method, in order to try to reduce the quenching errors for (ratios of) decay constants.
\section*{Acknowledgements}
We thank L.~Lellouch and S.~Sharpe for interesting discussions on the subject of this paper. We also thank F.~Parodi for informing us on the updated value of $f_{D_s}^{(exp.)}$.
D.B. acknowledges the support of ``La Fondation des Treilles''. V.L. and G.M. acknowledge  the M.U.R.S.T. and the INFN for partial support.
G.M. thanks the CERN TH Division for hospitality during the completion of this work. 
\newpage

\end{document}